\documentclass{iau} 
\usepackage{graphicx}

\def\be{\begin{equation}}
\def\ee{\end{equation}}

\def\Bcr{B_{\rm cr}}
\def\lsim{\lower 2pt \hbox{$\, \buildrel {\scriptstyle <}\over
         {\scriptstyle \sim}\,$}}
\newcommand\gsim{\buildrel > \over \sim}

\title[Ultra-Strong Magnetic Fields] 
{Physics in Ultra-Strong Magnetic Fields}

\author[A. K. Harding]   
{Alice K. Harding$^1$}

\affiliation{$^1$Theoretical Division, Los Alamos National Laboratory\\Los Alamos, NM 87545 USA  \\ email: {\tt ahardingx@yahoo.com}}

\pubyear{2022}
\volume{363}  
\jname{Neutron Star Astrophysics at the Crossroads:
Magnetars and the Multimessenger Revolution}
\editors{E. Troja \& M. Baring, eds.}
\begin{document}

\maketitle

\begin{abstract}
Several populations of neutron stars have surface magnetic fields above the critical strength of $4.4 \times 10^{13}$ G where the electron cyclotron energy equals its rest mass.  These include high-field rotation-powered pulsars, X-ray dim isolated neutron stars (XDIN), and magnetars.  In such ultra-strong fields, quantum effects in physical processes as well as additional exotic Quantum Electrodynamic processes only occurring at these high field strengths have a significant influence on the emitted radiation.  Although very strong magnetic fields play a critical role both inside and outside of neutron stars, I will review primarily processes that operate in the neutron star magnetospheres and how they influence the observed radiation.
\end{abstract}

\firstsection 
\section{Introduction}

Neutron stars (NS) have the strongest known magnetic fields in the Universe, with the highest exceeding $10^{15}$ G.  Although most of the field strengths are derived from the measured periods, $P$, and period derivatives, $\dot P$, assuming dipole spin down, they are likely within a factor of ten of the true magnetic fields.  At these extreme field values, physical process are strongly affected by Quantum Electrodynamic (QED) effects such as quantization of particle energies and birefringence of photon propagation modes.  Additional processes become possible, such as one-photon pair production and photon splitting, that cannot occur in field-free environments.  The field strength that marks the regime where these effects become important is the quantum critical field, $\Bcr = m^2 c^3 / (e \hbar) = 4.413 \times 10^{13}$ G, when the electron's cyclotron energy equals its rest mass energy.  However, QED effects can actually come into play around $B \sim 0.1 \Bcr$.  

\begin{figure}[t] 
\begin{center} \includegraphics[width=6.0in]{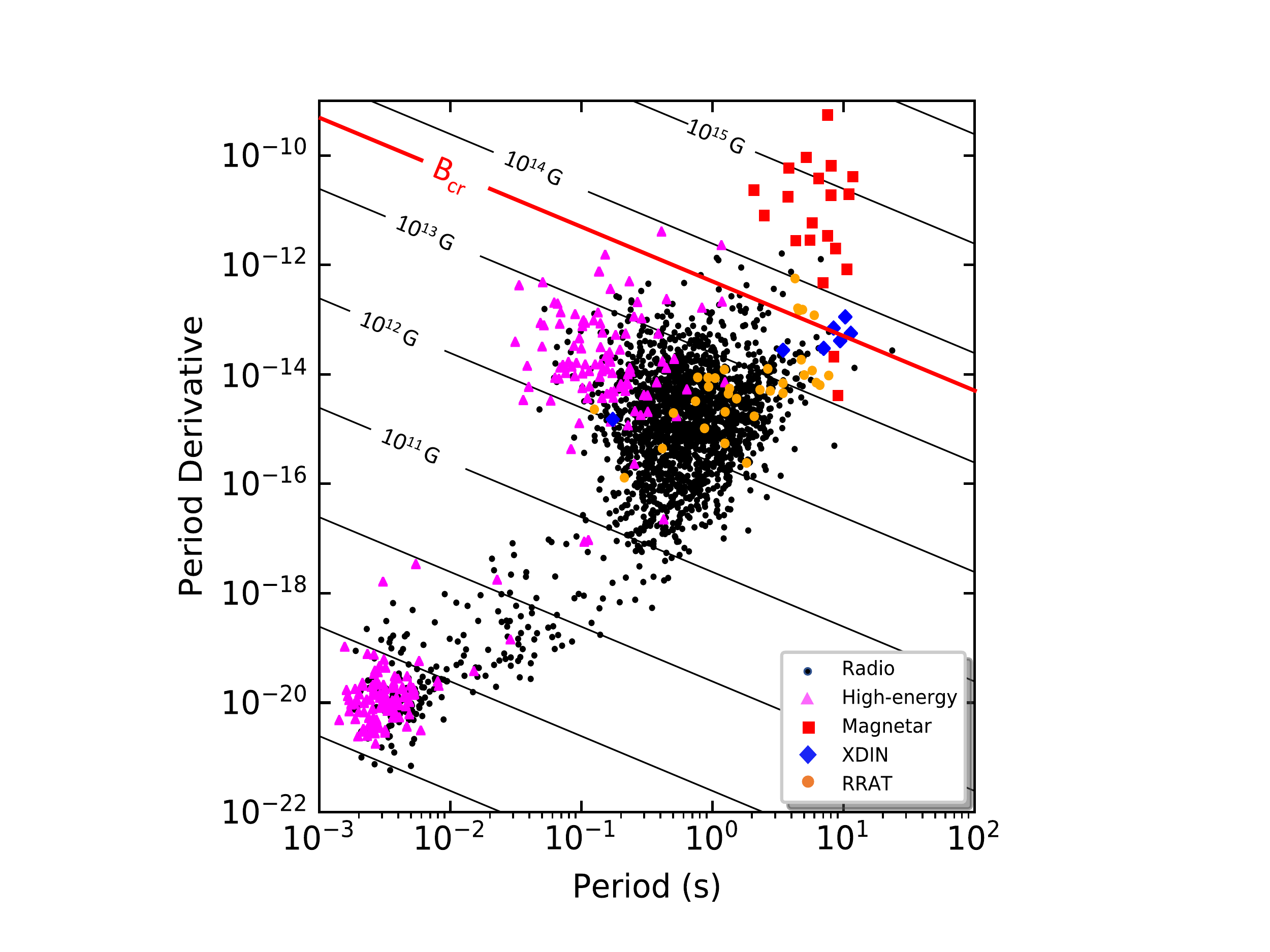} 
 \caption{Different types of neutron stars plotted on the period-period derivative diagram, with lines of constant surface magnetic field assuming dipole spin down. Data from https://www.atnf.csiro.au/research/pulsar/psrcat/.}
   \label{fig:PPdot}
\end{center}
\end{figure}

Figure \ref{fig:PPdot} shows some of the different types of NSs on the $P$-$\dot P$ diagram with lines of constant polar surface magnetic field $B_0 = 6.4 \times 10^{19} (P \dot P)^{1/2}$ G assuming dipole spin down, with $P$  in units of seconds and $\dot P$ in units of $10^{-15}\,s\,s^{-1}$.   The solid red line marks the value of $\Bcr$.  It is evident that all five NS types plotted, rotation-powered pulsars, magnetars, high-energy pulsars, XDINs and rotating radio transients (RRATs), have some members that lie above the $\Bcr$ line.  Interestingly, the different NS types tend to display very different behaviors even though they occupy the same part of the $P$-$\dot P$ diagram, indicating that derived dipole field strength cannot be the sole predictor of their true characteristics.  Since the observed $\dot P$ in most sources is measuring the dipole component of the magnetic field, unless very powerful winds are present (\cite{Harding1999}), it will be a lower limit on the true surface fields that may contain higher multipoles (\cite{Bilous2019,Kala2021}).  

This paper will present a brief review of physics in magnetic fields near and above $\Bcr$ that must be considered when modeling many of the sources in Figure \ref{fig:PPdot}.  For a more detailed discussion of these topics, refer to \cite{HardingLai2006}.

\section{Electron and Positrons States}

In strong fields, charged particle energies are quantized in eigenstates that are solutions to the Dirac Equation in a homogeneous magnetic field.   The energy states (or Landau states) are
\be
E_n  = (1  + p^2  + 2nB')^{1/2} ,  n = l + {\textstyle{1 \over 2}}(s + 1) = 0,1,2, \cdots 
\ee
where $n$, $l$, and $s$ are the principal, orbital and spin quantum numbers, respectively and 
$B' = B/B_{\rm cr}$, expressed in units with $c = \hbar =1$, and energies in $mc^2$.  Particles may have either spin up ($s=1$) or down ($s=-1$), except in the ground state $n=0$, where only the spin-down state is allowed.  The two spin states in levels $n>0$ are thus degenerate, but this degeneracy is broken by radiative corrections, allowing spin-flip transitions within Landau states (more on this below).  The momentum component parallel to the field, $p$, is continuous but momentum perpendicular to the field is quantized.  In magnetic fields $B' > 1$, the transverse motion becomes relativistic for electrons and positrons.  In the non-relativistic limit ($p \ll 1, B' \ll 1$), the transverse energy becomes $neB/mc$, multiples of the cyclotron energy. 

The above energy states and electron and positron wavefunction solutions in Cartesian coordinates were first presented by \cite[Johnson \& Lippmann (1949)]{JL1949} and are eigenstates of the kinetic momentum operator, $\pi = p + eA/c$.  Later, solutions in cylindrical coordinates were introduced by \cite[Sokolov \& Ternov (1968)]{ST1968} and are eigenstates of the magnetic moment parallel to $B$. While many derivations of physical processes in strong magnetic fields used the Johnson \& Lippmann (JL) wavefunctions, it was noted that the JL and ST wavefunctions have different spin-state dependences (\cite{Melrose1983}) and consequently will produce different spin-dependent results for physical processes, although the spin-averaged results agree.  Since the ST states have a number of desirable properties (\cite{Baring2005}), such as symmetry between electron and positron wavefunctions and the preservation of spin states under Lorentz transformations along $B$, it is the presently preferred formulation to use in derivation of high-magnetic field processes.  An additional important property of ST states is that the timescale for spin-flip transitions within Landau states is much slower than the decay timescale between Landau states (\cite{Geprags1994}), in contrast to JL states where such spin-flip timescales are comparable to those of Landau state transitions.   As will be discussed later in  \ref{sec:CompScat}, for scattering in the cyclotron resonance it is particularly important to use the cross section derived using ST states that treat the spin-dependent line widths correctly.

\section{Photons States}

Strongly magnetized plasmas are anisotropic and birefringent, so that photon polarization states and propagation are significantly influenced.  Vacuum polarization, in which a photon can convert to a virtual electron-positron pair which then annihilate, will dominate the photon propagation modes at soft X-ray energies when the density
\be
\rho_{_{\rm V}} < 1\,{\rm g\,cm^{-3}}\,\epsilon_{\rm keV}^2 B_{14}^2
\ee
(\cite{Lai2002}), where $\epsilon_{\rm keV}$ is the photon energy in units of keV and $B_{14}$ the magnetic field in units of $10^{14}$ G.  At higher densities above $\rho_{_{\rm V}}$, plasma modes will dominate the photon propagation so that a transition may occur between plasma and vacuum modes if a photon propagates across regions of changing magnetic field or plasma density.  Both plasma-dominated and vacuum-dominated polarization modes are nearly linear with electric vectors in (ordinary mode, $\parallel$ or O mode) or perpendicular to (extraordinary, $\perp$ or X mode) the plane of the photon momentum and magnetic field. For non-accreting NS sources, the plasma densities are low enough that vacuum modes dominate throughout the magnetosphere.

Although the full dielectric tensor for a magnetized plasma is quite complicated (\cite{Adler1971}), simplified expressions have been derived for low ($B \ll \Bcr$) and high ($B \gg \Bcr$) field limits of vacuum dispersion (\cite{Tsai1975}).   In both cases, the X and O mode indices of refraction both depend on the square of the angle between the photon propagation direction and the magnetic field, such that photons propagating along $B$ do not experience any vacuum dispersion.

\section{Radiative Processes in Strong Magnetic Fields}. \label{sec:RadProc}
\subsection{Cyclotron and synchrotron emission and absorption} \label{sec:synch}

\begin{figure}[t] 
\includegraphics[width=2.6in]{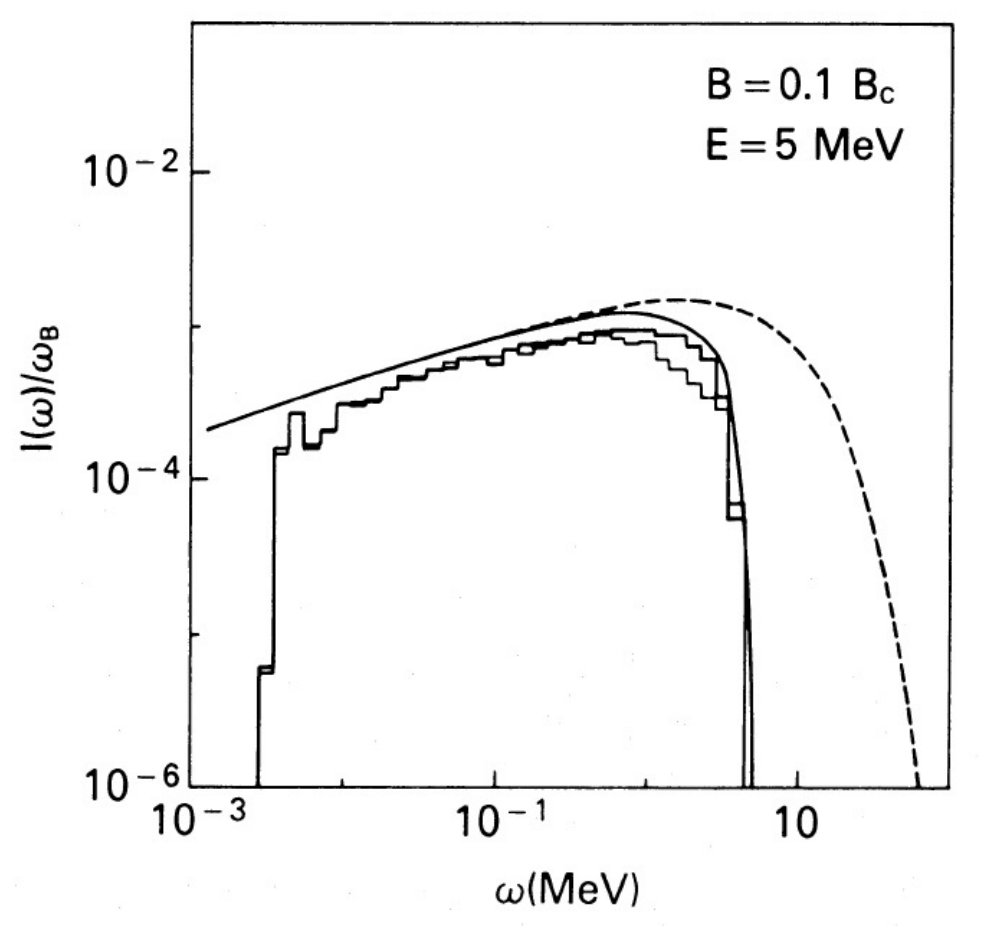} \includegraphics[width=2.6in]{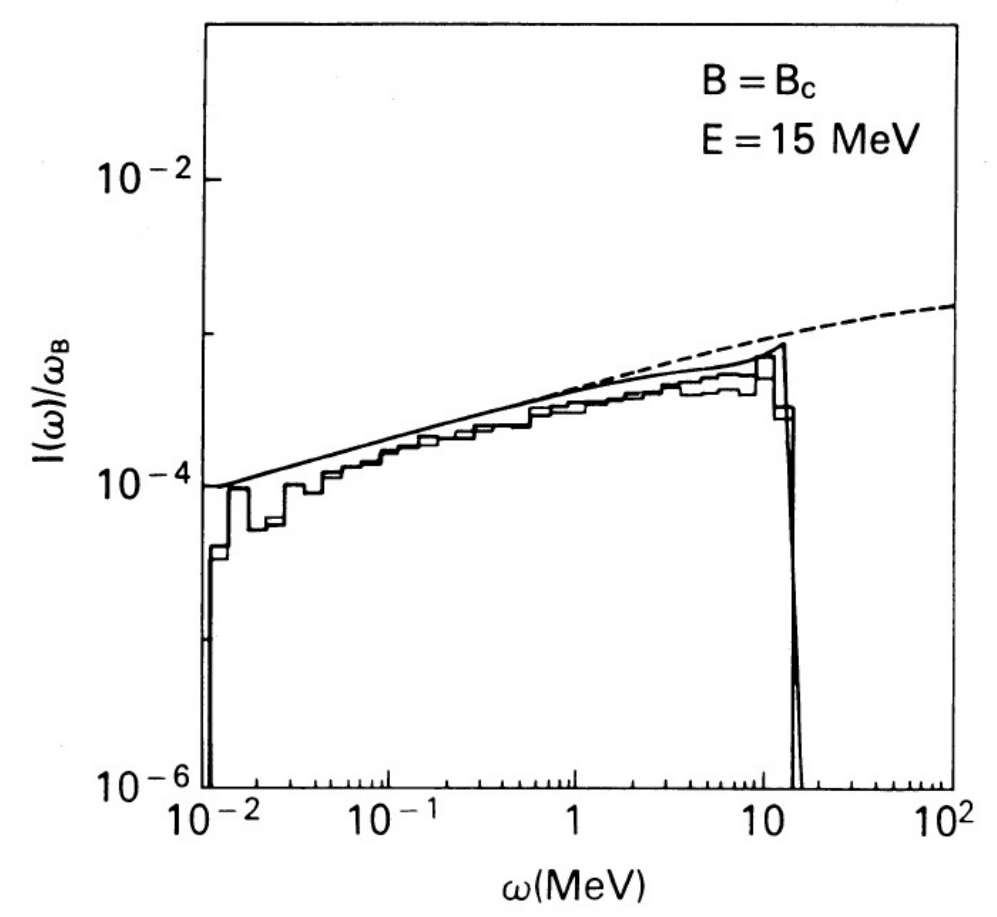} 
 \caption{Synchrotron emissivity (in units of the cyclotron frequency)  for single electrons with spin up (light-line histogram) or spin down (dark-line histogram).  The solid curve is the quantum asymptotic formula and the dashed curve is the classical formula. From \cite[Harding \& Preece  (1987)]{Harding1987}.}
   \label{fig:SR}
\end{figure}

At low field strengths, cyclotron radiation (and synchrotron radiation at relativistic energies) is a process in which a charged particle radiates as it spirals continuously around the magnetic field.  At high field strengths, the particles are confined to Landau states where their momentum perpendicular to the field is quantized, and the emission is described by transitions between states.  In fields $B \gsim 0.1 \Bcr$, the cyclotron energy, and thus the energies of the emitted photon, is a significant fraction of the particle rest mass and the particle recoil is important.  Cyclotron decay rates in the QED regime have been derived by \cite[Sokolov \& Ternov (1968)]{ST1968} for transitions between arbitrary Landau levels, spin states and photon polarizations.  The rates for spin-flip transitions are generally lower than those for transitions where the particle keeps the same spin.  For emission from very high Landau levels, $n \gg 1$, simple asymptotic formulae have been derived by \cite[Sokolov \& Ternov (1968)]{ST1968} that well-describe some of the peculiarities of cyclotron/synchrotron radiation in high magnetic fields.  One of these is the so-called dominance of ground-state transitions, which appears for $B' \sim 0.2$, where a particle in a high Landau state prefers to make a transition to the ground state radiating a photon of nearly its entire energy, rather than to an adjacent state as in classical synchrotron radiation.  Such ground state transitions produce a rise or tip at the high-energy end of the spectrum, which cuts of abruptly at the particle energy, as shown in the right panel of Figure \ref{fig:SR}.  On the other hand, the classical synchrotron radiation formula violates this conservation of energy when 
\be
\gamma ^2 B'\sin \psi  > \left( {\gamma  - 1} \right)
\ee
or when the classical critical frequency exceeds the particle energy, where $\psi$ is the pitch angle.  The classical synchrotron energy loss formula also overestimates the actual loss rate when $\Gamma \equiv \gamma B'\sin \psi   > 0.1$, there the energy loss rate is approximately $\dot\gamma \propto \gamma^{2/3}$ (\cite{Erber1966}), compared to the classical formula where $\dot\gamma \propto \gamma^2$. 

Cyclotron absorption, the inverse of cyclotron emission, is a first-order process in which a
photon excites a particle to a higher Landau state.  For particles in the ground state with $p = 0$, the 
required energy for excitation to state $n$ is (in rest mass units)
\be
\epsilon _n  = [(1 + 2nB'\sin ^2 \theta )^{1/2}  - 1]/\sin ^2 \theta 
\ee
for a photon propagating at angle $\theta$ to the field (\cite{Daugherty1978}).  Because of the recoil of the particle, the cyclotron harmonics are actually anharmonic in high magnetic fields, so that the energy difference between successive harmonics decreases.

\subsection{Compton scattering}  \label{sec:CompScat}

In strong magnetic fields, the cyclotron decay rate is much higher than any other processes that can de-excite Landau states, such as collisions, even for accreting NS sources.  So when an electron or positron undergoes cyclotron absorption to an excited state it always de-excites by cyclotron emission and this process is really Compton scattering.  Compton scattering is a second order process but becomes first order at the cyclotron resonances, where the intermediate virtual electron or positron is real.  In the non-relativistic (Thompson) limit, the scattering cross section has a single resonance at the fundamental (\cite{Canuto1971}), with the opacity for O mode much higher than for X mode below the resonance.  The relativistic (QED) scattering cross section (\cite{Daugherty1986,Sina1996}) has resonances at all the higher harmonics as photons can scatter the electrons to higher Landau states.  Transitions to higher states will then spawn more photons through one or more de-excitations.  

\begin{figure}[t] 
\begin{center} \includegraphics[width=3.0in]{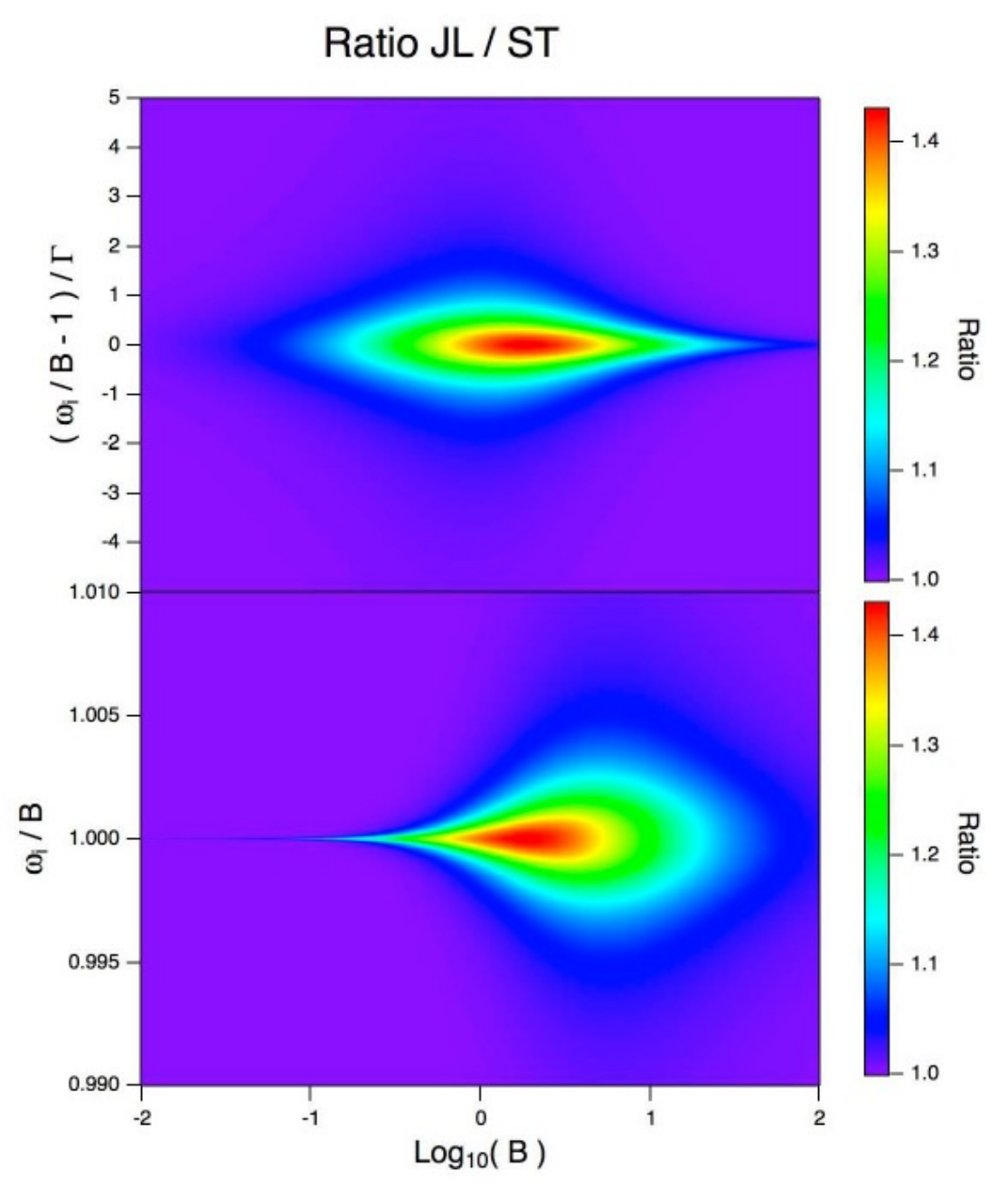} 
 \caption{Contour plots of the ratios of Johnson-Lippmann to Sokolov-Ternov Compton scattering cross sections as a function of $B$, over a range of photon energies that cover the cyclotron resonance ($\omega_i = B$), for photons propagating along $B$.  The upper panel displays the energy range as a ratio of natural line width $\Gamma$.  From \cite[Gonthier et al. (2014)]{Gonthier2014}. }
   \label{fig:JL/ST}
\end{center}
\end{figure}

At the cyclotron resonances, the scattering cross section will be infinite (since the denominator goes to zero for exact conservation of energy in the virtual state) unless one takes into account the natural line width, which is the decay rate of the Landau states of the intermediate particles.  This is relatively simple in the case of the magnetic Thompson cross section since only the fundamental resonance is present and there is no spin dependence, so one can add a term for the line width to the denominator (\cite{Daugherty1989}) to form a Lorentz profile.  In the case of the QED cross section, this solution is more complicated since the resonance in the cross section is a sum over all intermediate virtual states which may occupy many increasing Landau and spin states.  Thus the prescription for adding the resonant line widths becomes spin dependent and \cite[Graziani (1993)]{Graziani1993} showed that it is necessary to use eigenstates that diagonalize the magnetic moment operator parallel to $B$, i. e. ST states, not JL states.  \cite[Gonthier et al. (2014)]{Gonthier2014} demonstrated that using JL states will overestimate the resonant cross section including the spin-dependent line widths compared to using the correct ST states for the case of photons propagating along $B$, as shown in Figure \ref{fig:JL/ST}.  The full scattering cross section with natural line width included in ST formalism was presented by \cite[Mushtukov et al. (2016)]{Mushtukov2016}.

\begin{figure}[t] 
\begin{center} \includegraphics[width=5.0in]{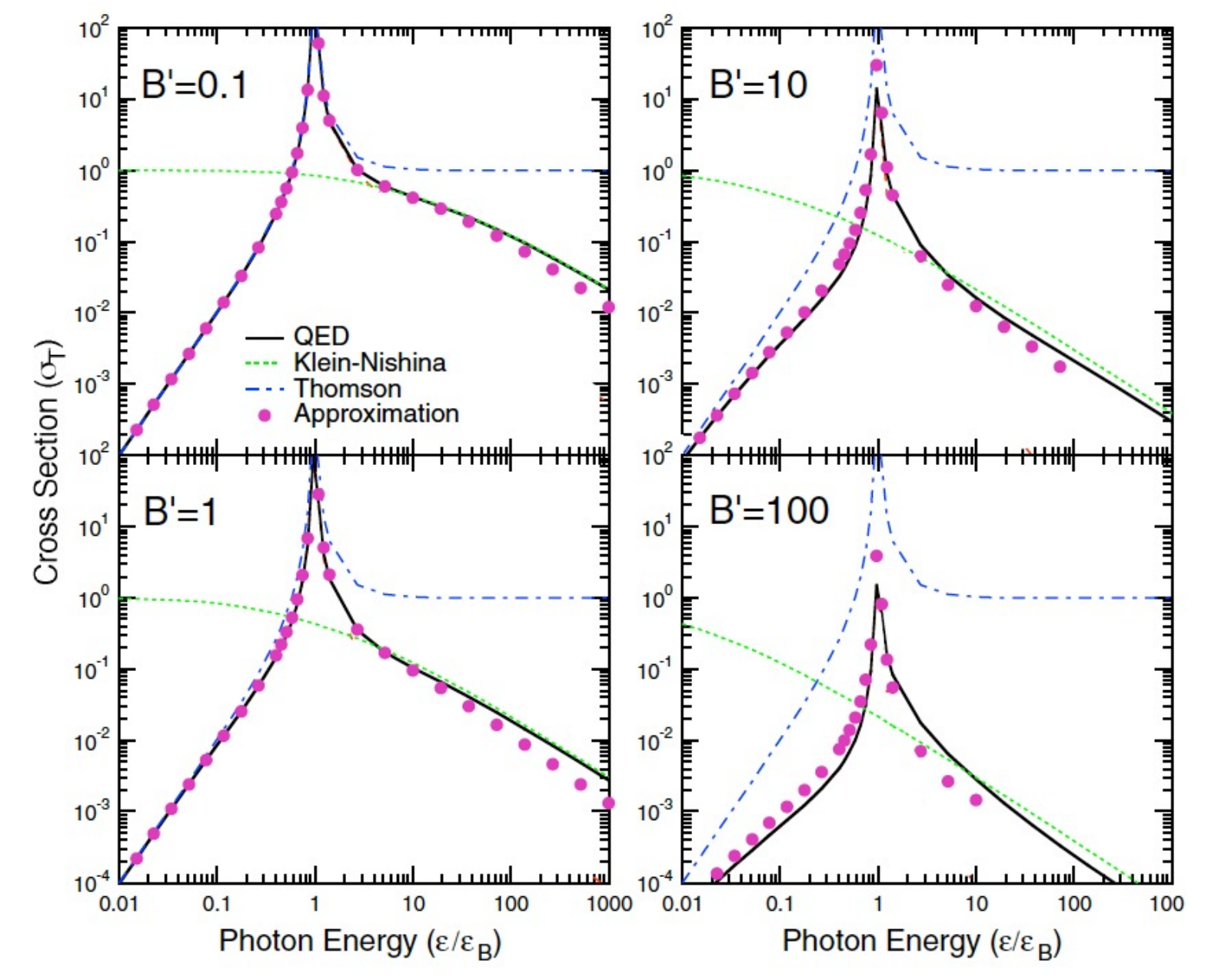} 
 \caption{Total polarization-averaged Compton scattering cross section for an electron at rest in
the ground state, in units of the Thompson cross section, as a function of incident photon energy
in units of cyclotron energy, for the case of incident photon angle $\sin\theta = 0$. Solid line: relativistic
QED cross section (e.g. \cite[Daugherty and Harding (1986)]{Daugherty1986}) summed over final
electron Landau states up to 500, dot–dashed line: magnetic Thompson cross section,
dotted line: Klein-Nishina (non-magnetic) cross section, dots: approximate QED cross section from \cite[Gonthier et al. (2000)]{Gonthier2000}. 
 Adapted from \cite[Gonthier et al (2000)]{Gonthier2000}.}
   \label{fig:Gonthier2000}
\end{center}
\end{figure}

The full scattering cross section however is quite cumbersome to use in most modeling.  Fortunately some simplified expressions have been developed for specific cases.  A useful expression for the case of relativistic electrons ($\gamma \gsim 10$) moving along $B$, that `see' the incident photons propagating head on along $B$ in their rest frames, was presented by \cite[Gonthier et al. (2000)]{Gonthier2000}, showing that the QED cross section should be used for $B' \gsim 0.1$ (see Figure \ref{fig:Gonthier2000}).  In the case of scattering at the resonance, simplified expressions have been provided by \cite[Nobili et al. (2008b)]{Nobili2008b}\footnote{But note that JL states are used there.}, \cite[Gonthier et al. (2014)]{Gonthier2014} and \cite[Mushtukov et al. (2016)]{Mustukov2016}.  Finally, simplified expressions for QED cooling rates are given by \cite[Baring et al. (2011)]{Baring2011}.

Since Compton scattering at the resonance is first order, when can scattering be approximated as cyclotron absorption and emission?  In the case of the total cross section, the absorption  approximation is fairly accurate up to $B' \sim 0.1$ (\cite{Harding1991}).  However, the scattered photon distribution depends on the differential cross section which can be significantly different from that of absorption at even field strengths $B' < 0.1$ (\cite{Nobili2008b}).

\section{Photon Attenuation Processes}  \label{sec:atten}

Since strong magnetic fields are able to absorb or supply momentum perpendicular to $B$, only momentum parallel to $B$ is conserved in processes involving transitions between Landau states.  This enables a number of additional processes to occur that cannot conserve energy and momentum in field-free space.  For example, two photons can create an electron-positron pair in a field-free environment but a single photon cannot.  However, a strong field can absorb the extra momentum of the photon with enough energy to create a pair.  

\subsection{One-photon pair production}

One-photon (or magnetic) pair production is a first order QED process in which a single photon with energy above the threshold, $\varepsilon_{\rm th} = 2/\sin\theta_{\rm kB}$, converts to an electron-positron pair, where $\theta_{\rm kB}$ is the angle between the photon momentum and $B$.   At fields $B' \ll 0.1$, the attenuation coefficient increases exponentially as $\exp{(-4/3\chi)}$, where $\chi = \varepsilon B' \sin\theta_{\rm kB}/2$ (\cite{Erber1966}).  Therefore, pair production is probable when $\chi \gsim 0.1$.  However, this asymptotic expression is valid only well above threshold and for field strengths $B' > 0.1$ pairs are produced near threshold in low Landau states (\cite{Daugherty1983}), where the attenuation coefficient displays resonances at photon energies at which additional pair states become kinematically accessible.  For these higher field strengths, the exact QED cross section should be used but fortunately only the first few pair states need be included and the expressions for these are very simple (\cite{Baring2001}).  The one-photon pair attenuation rate is polarization dependent with $\parallel$ mode photons producing pairs in the ground Landau state while the threshold for $\perp$ mode photons is the first excited state.  Therefore in very strong fields, pairs produced by $\parallel$ mode photons in the ground state cannot radiate any cyclotron photons and pairs produced by $\perp$ mode photons can produce only photons at the cyclotron resonance, strongly suppressing pair cascades.

\subsection{Photon splitting} \label{sec:split}

A strong magnetic field allows a single photon to directly create two lower-energy photons, whereas this process is forbidden in free space by Furry's theorem (\cite{Baring1998}).  Since it is a third order process, the cross section is well below that of one-photon pair production but there is no threshold, so photon splitting will be important below pair threshold.  The attenuation rate, $\propto \varepsilon^5 (B' \sin\theta_{\rm kB})^6$, is a strong function of both $B$, angle and photon energy (\cite{Adler1971}).  Since radiating particles in very high magnetic fields travel parallel to magnetic field lines, the high-energy photons are emitted with very small angles to $B$.  Therefore both pair and photon splitting attenuation rates will initially be zero and neither process will occur until the photons propagate some distance across curved magnetic field lines.  Photon splitting can thus prevent pair production if photons can split before they reach pair threshold, and studies in dipole fields show that this happens for magnetic field strengths $B' \gsim 1$ (\cite{Baring1998,Baring2001}).  Simplified expressions for the attenuation lengths and escape energies for both photon splitting and one-photon pair production are given in \cite[Hu et al. (2019)]{Hu2019}.

Both pair production and photon splitting rates depend on photon polarization and, while this behavior is well known for pair creation, the number of allowed splitting modes is currently unresolved.  Although there are three kinematically allowed splitting modes, $\perp \rightarrow \parallel\parallel$, $\perp \rightarrow \perp\perp$, $\parallel \rightarrow \perp\parallel$, \cite[Adler (1971)]{Adler1971} determined that only the first was allowed in the weakly dispersive limit.  However, in ultra-strong fields with high vacuum dispersion, and high plasma dispersion near the cyclotron resonance, the other modes may operate if higher-order non-linear contributions to the dispersion are significant.  Which modes are allowed is critically important for emission models, since photon splitting of both photon polarization modes will completely suppress pair creation and produce pure splitting cascades whereas splitting of only $\perp$ mode photons allows pair creation by $\parallel$ mode photons (\cite{Baring2001}).

\begin{figure}[t] 
\begin{center} \includegraphics[width=4.0in]{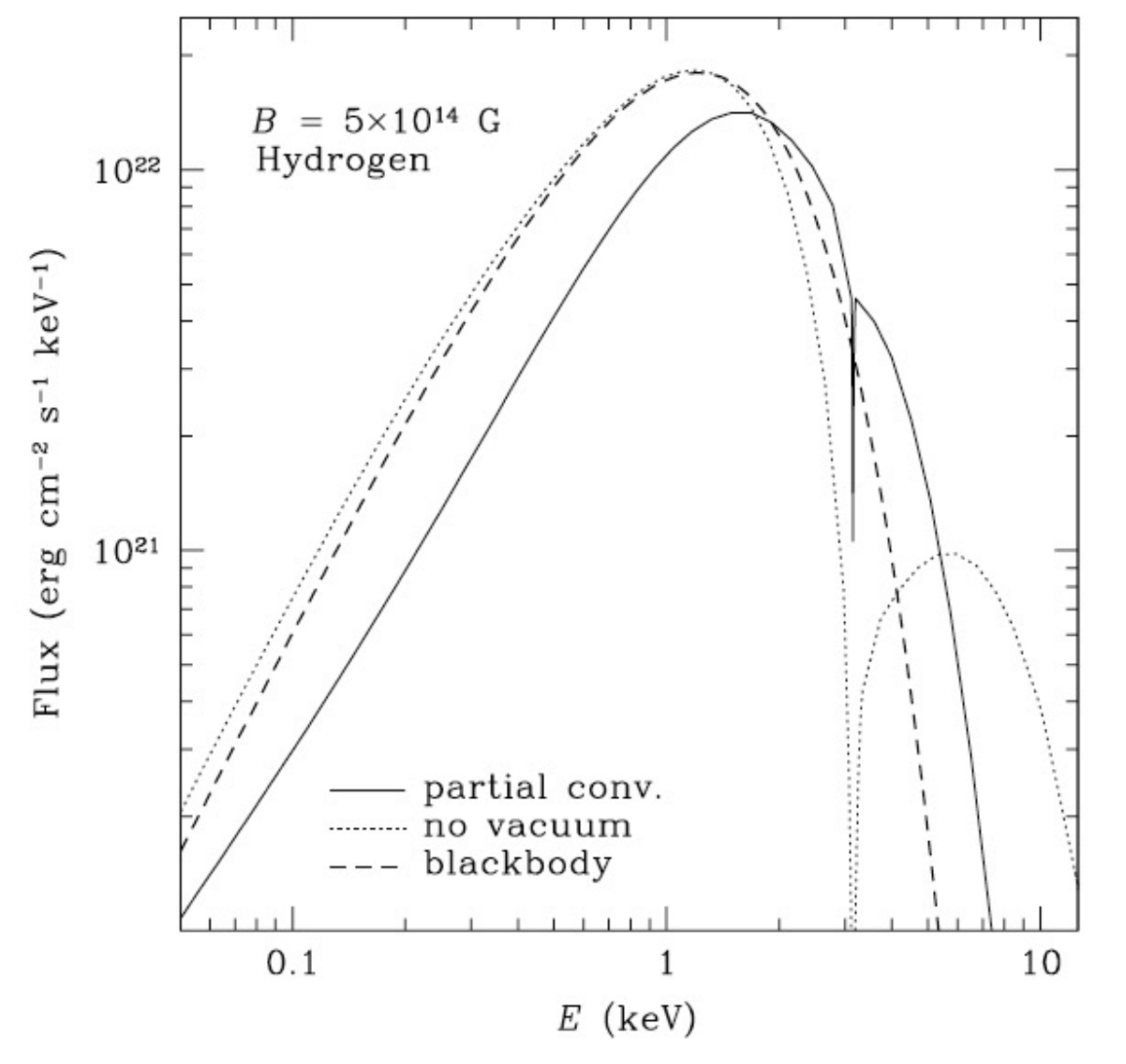} 
 \caption{Model spectrum for fully ionized atmosphere with $B = 5 \times 10^{14}$ G and $kT = 0.44$ keV, showing the effect of including vacuum polarization. From \cite[van Adelsberg \& Lai (2006)] {VanAdels2006}.}
   \label{fig:Van2006}
\end{center}
\end{figure}

\section{Radiation by Magnetars and Other High-Field Neutron Stars}

This section will highlight a selection of results using the processes discussed in Section\ref{sec:RadProc} and \ref{sec:atten} from models of emission and propagation in strong magnetic fields.  

\subsection{Strongly magnetized atmospheres}

Thermal emission has been detected from isolated NSs such as rotation-powered pulsars, magnetars and XDINs since the early 1980's.   Following the realization that Anomalous X-ray pulsars (AXPs) and Soft Gamma-Ray Repeaters (SGRs) have very high spin-down rates (\cite{Kouvel1998}) and thus extremely high magnetic fields, strongly magnetized NS atmospheres have been modeled over the past twenty years by a number of groups (\cite{Ozel2001,Ho2001,Zane2001,Barchas2021}).   Since these atmospheres are a thin layer of highly magnetized plasma above the NS surface, the transfer of the thermal photons emitted from the hot NS surface is both anisotropic and birefringent, so processes such as polarization-dependent scattering and vacuum polarization must be included and properly treated.  However, the atmospheric temperatures are low enough, $kT \sim 1$ keV, that the photon energies are well below the electron cyclotron frequency in magnetar fields and non-relativistic, so that use of QED scattering cross sections are not necessary.  Because the X and O mode opacities below the cyclotron frequency are very different, the radiation emerging from highly magnetized atmospheres is strongly polarized in the X mode.

Most models for magnetars treat ionized H and He atmospheres and assume an underlying NS temperature distribution, which is not really known {\it a priori}.  Scattering at the ion cyclotron resonance, at several keV, will produce absorption-like features in the spectrum.  In conventional NS magnetic fields ($B \sim 10^{12} - 10^{13}$ G) the vacuum resonance lies outside the photospheres of both O and X mode photons.
However in magnetar fields, the vacuum resonance lies between the two photospheres and conversion between X and O modes can strongly affect the photon transfer and suppress the ion cyclotron features (\cite{Ho2003}) (see Figure \ref{fig:Van2006}).

\subsection{Photon transport in neutron star magnetospheres}

The photons that emerge from the NS atmospheres must still propagate through the magnetosphere which for magnetars is dominated by vacuum birefringence.  \cite[Heyl et al. (2003)]{Heyl2003} argue that in a magnetized vacuum surrounding a NS, the polarization modes will propagate adiabatically and  independently if the magnetic field changes slowly enough and is high enough that the indices of refraction of the X and O vacuum modes are very different.  In this case, the polarization directions will follow the magnetic field direction out to a polarization-limiting radius which if far enough from the NS can probe the field structure and preserve the high polarization of the atmosphere radiation.

Studies of polarized transfer through highly-magnetized magnetospheres, assuming input of photons from either magnetized atmospheres or condensed surfaces, has been carried out by a number of groups.  \cite[Gonzalez Caniulef et al. (2016)]{Gonzalez2016} showed that polarized transport of photons from both an atmosphere and condensed surface through a strongly magnetized dipole magnetosphere preserved the much higher initial polarization of the atmosphere.   Thus, observing very high degrees of polarization in magnetar thermal emission ($\gsim 50\%$) will signal vacuum birefringence.   However, lower levels of polarizations will not be conclusive since photons from a condensed surface have a lower level of intrinsic polarization.    \cite[Fernandez \& Davis (2011)]{Fernandez2011} studied the transport of blackbody photons polarized 100\% in X mode through a magnetosphere with twisted magnetic fields, including vacuum polarization and resonant scattering, finding that the emergent degree of polarization increases with the amount of field twist.  

\subsection{Resonant Inverse Compton scattering}

\begin{figure}[t] 
\begin{center} \includegraphics[width=5.5in]{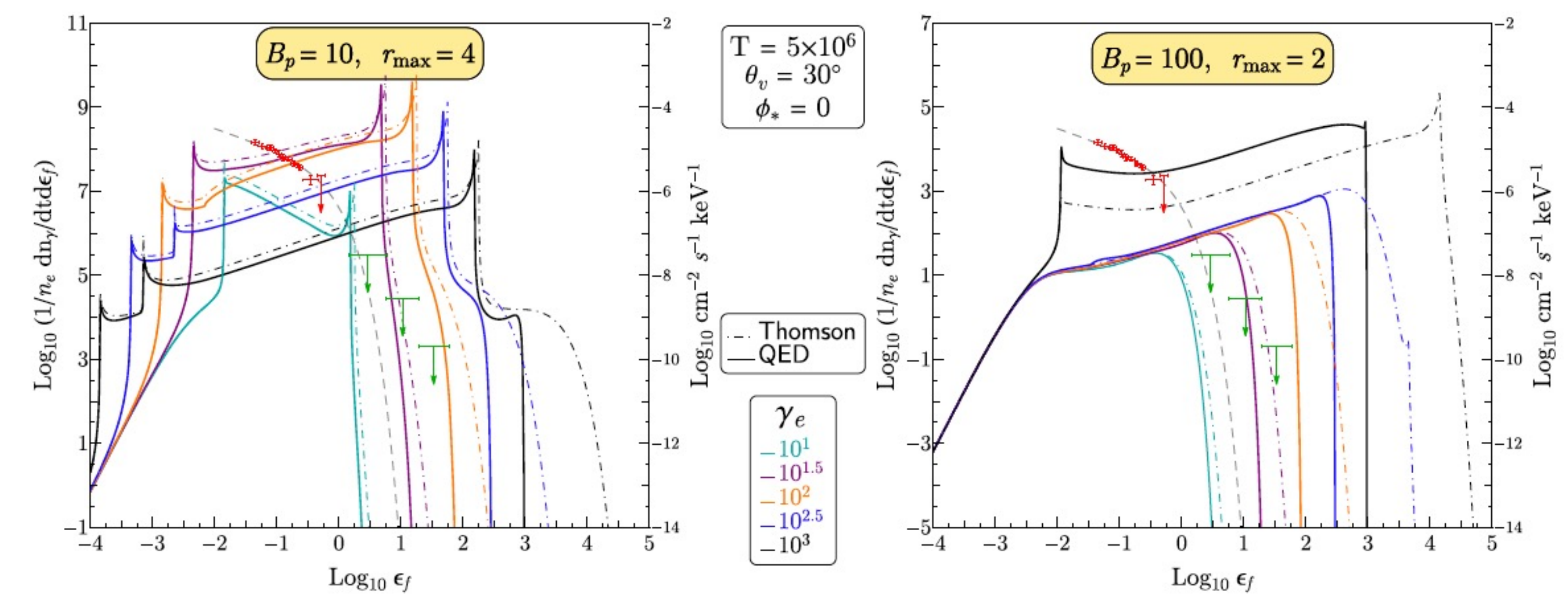} 
 \caption{Model resonant inverse Compton scattered spectra from uncooled, mono-energatic particles of Lorentz factor $\gamma_e$ scattering thermal photons with temperature $T = 5 \times 10^6$ K on closed dipole field loops that extend to maximum radius $r_{\rm max}$ from the NS for observer viewing angle $\theta_v$. From \cite[Wadiasingh et al. (2018)]{Wadias2018}. }
   \label{fig:Wad2018}
\end{center}
\end{figure}

The hard tails seen in many magnetars extending from $\sim 10$ keV to at least 200 keV have been modeled as resonant Inverse Compton scattering (RICS) of thermal surface emission by mildly relativistic electrons (\cite{Baring2007,Nobili2008a,Belo2013}).
Compton scattering in the high fields of magnetar magnetospheres for scattered photon energies above $\sim 10 - 20$ keV requires use of QED cross sections or one of the approximate descriptions discussed in Section \ref{sec:CompScat}.  Modeling of scattered spectra depend on the angular distribution of final photons which are significantly affected by QED effects above 10 keV (\cite{Nobili2008b}).  Additionally, in modeling resonant scattering the use of cross sections or approximations derived using the ST states is critical.  Figure \ref{fig:Wad2018} shows an example of RICS spectra modeled using both the non-relativistic magnetic Thompson and QED cross sections (using approximate expressions of \cite[Gonthier et al. (2014)]{Gonthier2014} and \cite[Baring et al. (2011)]{Baring2011}) for different particle Lorentz factors (\cite{Wadias2018}).  It is apparent that the Thompson cross section over-estimates the high-energy spectral cutoff, predicting scattered photons above the particle energy, with the discrepancy increasing with $B$ and $\gamma_e$, as the QED cross section includes Klein-Nishina reductions and electron recoil.

\subsection{Pair and photon splitting attenuation and polarization signatures}

\begin{figure}[t] 
\hskip -0.2cm
 \includegraphics[width=5.5in]{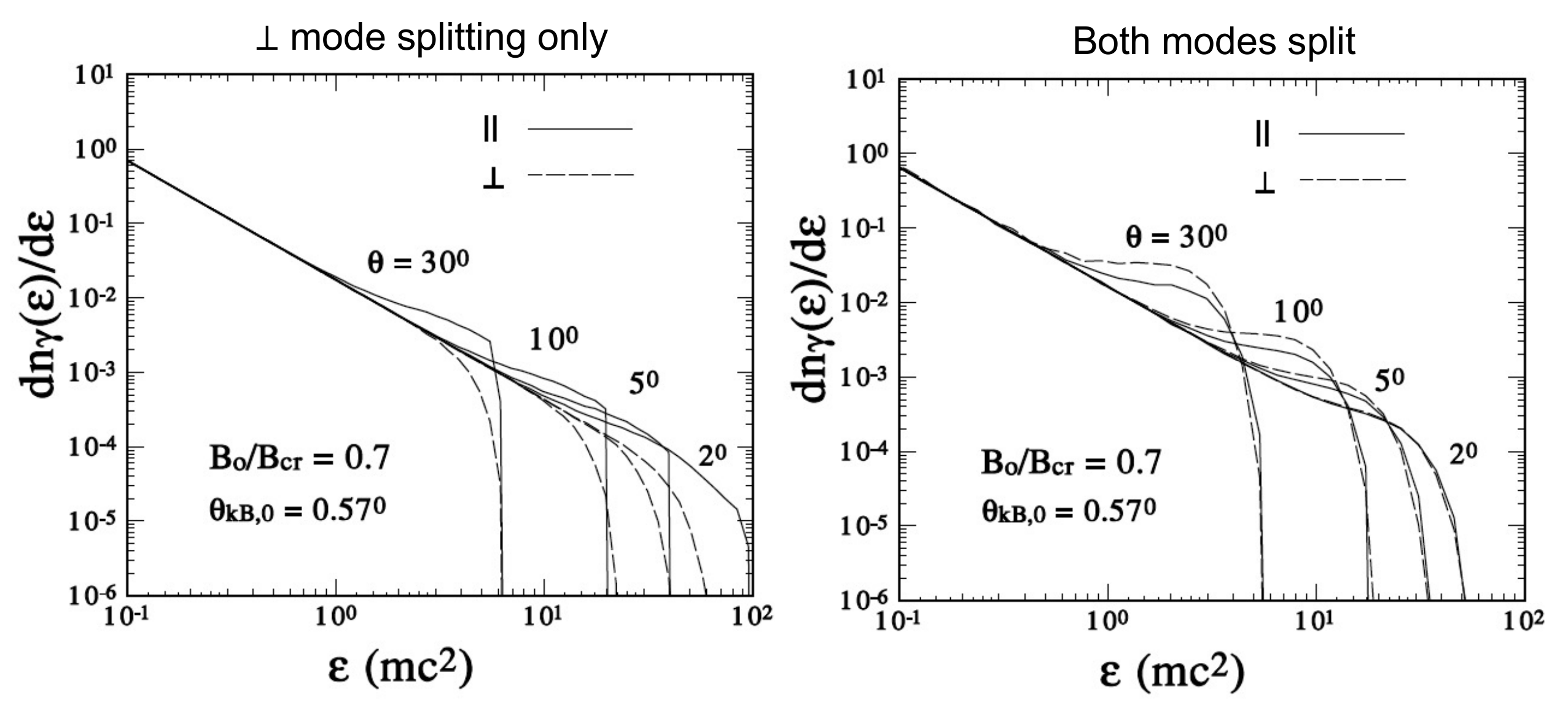} 
 \caption{Unpolarized power-law emission (with photon index -1.6) emitted at angle $\theta_{\rm kB} = 0.57^\circ$ to the
magnetic field at different magnetic colatitudes, $\theta$, as labeled, attenuated by pair production and photon splitting.  
 Left panel: photon splitting in only $\perp$ mode.  Right panel: photon splitting in both polarization modes.  The normalization of the spectrum is arbitrary From \cite[Harding et al. (1997)]{Harding1997}}
   \label{fig:Hard1997}
\end{figure}

One-photon pair production and photon splitting will significantly attenuate spectra at high energies, depending on the location of emission and observer viewing angle.  The strong polarization dependence of both process can also produce distinctive signatures near the spectral cutoffs.  Figure \ref{fig:Hard1997} illustrates the different predicted properties of the attenuation by pair production and photon splitting if only one mode (only $\perp$ mode photons split) or three modes (both $\perp$ and $\parallel$ mode photons split) are allowed (see Section \ref{sec:split}).  In the case where only one splitting mode is allowed, $\perp$ (X) mode photons split below pair threshold while $\parallel$ (O) mode photons split at the higher-energy pair threshold, so that the spectrum is nearly 100\% polarization in O mode before the cutoff.  In the case where all modes of splitting can occur, both X and O mode photons split below pair threshold and can produce a splitting cascade that hardens the spectrum that will be dominated by X mode below the cutoff.  Thus, although the study in Figure \ref{fig:Hard1997} is a model for the high-field pulsar B1509-58, it illustrates that measuring the polarization of magnetar hard spectra at the cutoff can detect photon splitting and distinguish which modes are operating.

\begin{figure}[t] 
\hskip -0.5cm
\includegraphics[width=6.2in]{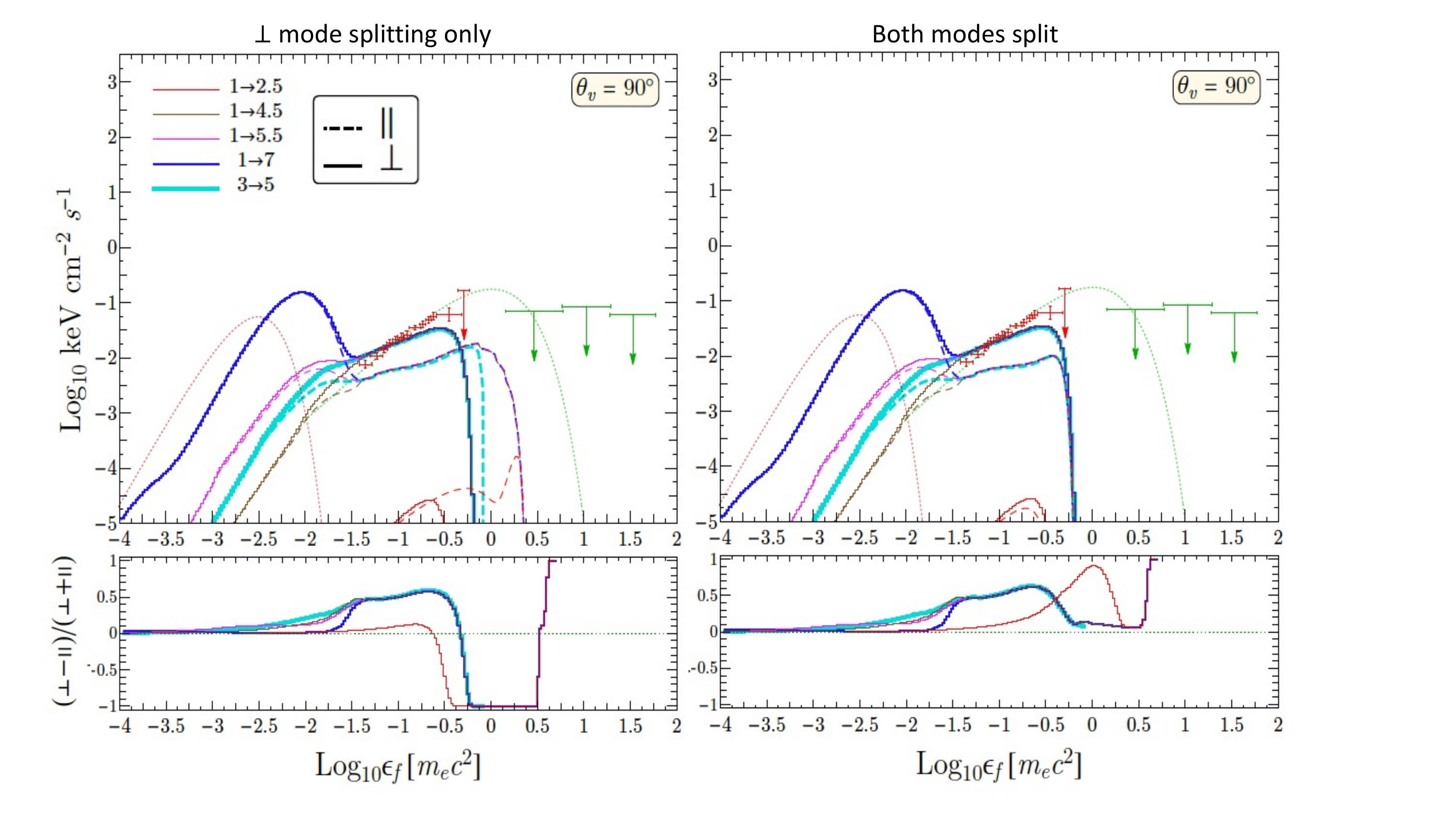} 
 \caption{Model resonant inverse Compton scattered spectra from uncooled, mono-energatic particles of Lorentz factor $\gamma_e = 10$ scattering thermal photons with temperature $T = 5 \times 10^6$ K for surface magnetic field $B' = 10$.  Emission has been integrated over ranges of closed dipole field loops with minimum and maximum radii labeled in the colors shown.  From \cite[Wadiasingh et al. (2022)]{Wadias2022}.}
   \label{fig:Wad2022}
\end{figure}

The examples shown in Figure \ref{fig:Hard1997} are for initially unpolarized photon spectra.  However, most radiation processes in high magnetic fields will produce polarized spectra and this will affect the signatures discussed above.  RICS produces spectra polarized in X mode.  However, \cite[Wadiasingh et al. (2022)]{Wadias2022} have shown that polarization signatures that can distinguish between splitting in one or three modes can still be observed.  The examples in Figure \ref{fig:Wad2022} display RICS spectra attenuated by pair production and photon splitting in one or both photon  modes and the polarization as a function of energy.   In the case of splitting in one (X) mode most of the RICS spectra that are polarized mostly in X mode are attenuated by splitting below pair threshold, leaving a lower but still potentially detectable level of pure O mode near the pair cutoff with a sharp change from X to O mode at the splitting cutoff.  In the case of splitting in both modes, the whole spectrum is attenuated by splitting below pair threshold so that the polarization stays in X mode up to the cutoff.
A half dozen magnetars have hard tail emission that is bright enough to be detected by proposed telescopes such as AMEGO-X (\cite{Fleis2021}) or COSI (\cite{Tomsick2021}) sensitive at MeV energies, making possible the detection of these predicted signatures of photon splitting (\cite{Wadias2019}).

\section{Conclusions}

With such a large population of detected neutron stars having inferred surface magnetic fields near and above the critical field, the use of QED physics to describe their emission has become more vital.  This review has attempted to provide a guide to the important processes and their properties, when it is necessary to consider QED effects in modeling these sources and when and which approximations may be used.  Vacuum polarization has been shown to be of importance for transfer of photons with energies well below $mc^2$ in magnetic fields below $\Bcr$, even though the non-relativistic magnetic Thompson scattering cross section may be safely used.  Scattering of photons at energies above $\sim 10-20$ keV should include QED effects, using either the full QED cross section of one of the appropriate approximations that are available.  Resonant scattering models in magnetic fields $B \gsim 0.1 \Bcr$ should use cross sections derived using ST eigenstates that correctly treat the intermediate spin state line widths.  

We are entering an era where detection of QED effects in emission from strongly magnetized NSs is possible.  Evidence for vacuum polarization has possibly already been detected in optical polarization of the XDIN source RX J1856.5-3754 (\cite{Mignani2017}).  Such polarization measurements of these sources and magnetars with the recently launched X-ray polarimeter IXPE (\cite{Weisskopf2021}) will offer an additional possibility for detection of vacuum polarization.  Finally with more sensitive MeV telescopes on the horizon is the exciting possibility of detecting photon splitting polarization signatures for the first time.

I would like to thank the editor, Matthew Baring, and Zorawar Wadiasingh for their review and suggestions.

\end{document}